\documentclass[10pt]{article}
\usepackage[cp866]{inputenc}
\usepackage[russian]{babel}

\begin{document}


\newcommand{\N}{N\raise.7ex\hbox{\underline{$\circ $}}$\;$}

\begin{center}
{\bf

\thispagestyle{empty}

BELARUS NATIONAL ACADEMY OF SCIENCES

B.I. STEPANOV's INSTITUTE OF PHYSICS

}
\end{center}

\vspace{20mm}

\begin{center}
{\bf
Red'kov V.M.\footnote{E-mail: redkov@dragon.bas-net.by}}

\end{center}

\vspace{10mm}

\begin{center}
{\bf
MONOPOLE  BPS-SOLUTIONS  OF THE  YANG-MILLS  EQUATIONS IN SPACE OF  EUCLID, RIEMANN,
and LOBACHEVSKI
}

\end{center}

\begin{center}
{\bf Abstract}
\end{center}

\begin{quotation}

Procedure of finding of  the  Bogomolny-Prasad-Sommerfield  monopole  solutions
in  the Georgi-Glashow   model  is  investigated  in  detail  on  the backgrounds
of  three  space models  of constant curvature:  Euclid,  Riemann,  Lobachevski's.
Classification of  possible solutions is given. It is shown that among all solutions
there exist  just  three  ones which reasonably and in  a one-to-one  correspondence
can be associated   with  respective   geometries.  It is pointed out that the known
non-singular BPS-solution in the flat Minkowski space  can be understood as a result
 of somewhat artificial combining the Minkowski space background with  a possibility
 naturally  linked up with the Lobachewski geometry. The standpoint is brought forth
that  of primary interest should be regarded only three specifically  distinctive
solutions --- one for every curved space background. In the framework of those
arguments  the generally accepted status of the known monopole BPS-solution
should be critically reconsidered and even might be given away.

\end{quotation}

\vspace{5mm}

{\bf PASC: 1130, 0230, 0365, 2110H}

\vspace{5mm}

{\bf Key Words:}   monopole, BPS-limit, non-Euclide geometry,
constant curvature,

\vspace{10mm}

7 pages, 17 references
\newpage

\subsection*{1. Radial equations}

In the literature, a SU(2)-monopole problem in the limit of
Bogomolny-Prasad-Sommerfield and on a curved space-time background
has attracted some interest [1-17]. Additional analysis of this system is performed
in the present work.
In a space-time with a metric tensor $g_{\alpha \beta}(x)$ let us consider
the Higgs isotriplet of scalar fields. Lagrangian of that system is given by
$$
L = {1 \over 2} \;g_{\alpha \beta}(x) \; (\partial _{\alpha} \Phi ^{a})
(\partial _{\beta} \Phi ^{a}) - {\lambda \over 4} \; (\Phi^{2} -
F^{2})^{2} .
\eqno(1.1)
$$

\noindent Clearly, covariant derivative of a scalar field  $\nabla _{\alpha} \Phi $
coincides with the ordinary derivative   $\partial_{\alpha} \Phi $.
The fields $\Phi^{a}(x)\; (a =1,2,3)$ are supposed to be real; correspondingly,
the Lagrangian is invariant under $SO(3.R)$ group transformations.
Making the symmetry transformations local, depended on coordinates $x^{\alpha}$,
we are to replace the $\nabla_{\alpha}$ by the gauge-covariant one
$$
\partial_{\alpha} \;\; \Rightarrow \;\; D_{\alpha} \Phi^{a} =
\partial_{\alpha} \; \Phi^{a}  +  e \; \epsilon _{abc} \;W_{\alpha}^{b} \;
\Phi^{c}  \;  .
\eqno(1.2)
$$

\noindent Here and in the following, isotriplet indices are used equally as
upper and lower ones.   In eq.(1.2)
the symbol $W_{\alpha}^{b}$ stands for the Yang-Mills isotriplet.
Gauge-invariant antisymmetric tensor has a form
$$
F^{a}_{\alpha \beta} =  \partial_{\alpha}\; W^{a}_{\beta}  -
\partial_{\beta}  \; W^{a}_{\alpha}  +  e \;\epsilon _{abc} \;
W_{\alpha}^{b} \; W_{\beta}^{c}   \;   .
\eqno(1.3)
$$

\noindent   Modifying  derivatives $\partial_{\alpha}$
in the  Lagrangian (1.1)  in accordance with  (1.2), and adding  the Lagrangian of free
Yang-Mills triplet, one can produce a complete Lagrangian   under consideration
$$
L = {1 \over 2} \; g_{\alpha \beta}(x)  D_{\alpha} \Phi ^{a}
 D _{\beta} \Phi ^{a}  -
 {\lambda \over 4}
(\Phi^{2} -  F^{2})^{2} -
 {1\over 4} g^{\alpha \rho}(x) g^{\beta\sigma}(x)
F^{a}_{\alpha \beta}\; F^{a}_{\rho \sigma} .
\eqno(1.4)
$$

\noindent From (1.4), in accordance with the variational principle,
one can derive equations
$$
{1 \over \sqrt{-g}} D_{\alpha} \sqrt{-g} D^{\alpha} \Phi^{a} =
-\lambda ( \Phi^{2} - F^{2}) \; \Phi^{a}  , \;
$$
$$
{1 \over \sqrt{-g}} D_{\alpha} \sqrt{-g}  F^{\alpha \beta}_{a} =
-e \epsilon _{abc} \Phi^{b} D^{\beta} \Phi^{c}  .
\eqno(1.5)
$$

In the following, all analysis will be done  for three (curved) space models:
 Euclid's --- $E_{3}$, Riemann's --- $S_{3}$, and Lobachevski's --- $H_{3}$.
At this conformally flat coordinates  will be used:
$$
dS^{2} =  (dx^{0})^{2} +  {1 \over \Sigma^{2} } [
(dx^{1})^{2}  +   (dx^{2})^{2}   +   (dx^{3})^{2}]   )  , \;
$$
$$
r = \sqrt{(x^{1})^{2} +  (x^{2})^{2}+  (x^{3})^{2} } \; .
\eqno(1.6)
$$

\noindent To $E_{3}$-model there corresponds  $\Sigma  = 1$, to
$S_{3}$- model --- $\Sigma = (1 + r^{2} /4 )$, and for $H_{3}$-model ---
$\Sigma  = (1 - r^{2} /4 )$.

Let us be solving eqs. (1.5) in spaces (1.6)  with the use of the known
Julia-Zee (dyon) ansatz:
$$
\Phi^{a} = x^{a}  \Phi (r)  ,  \qquad  W^{a}_{0} = x^{a}  f(r)  \; ,
\eqno(1.7)
$$

\noindent
After making all necessary calculation one can reach at radial equations for
$\Phi (r),  f(r), K(r)$:
$$
\Phi''  +  {4\over r}  \; \Phi' - 2 e\; \Phi \;
(2 + e r^{2}K)  \; K -   {\Sigma ' \over  \Sigma} \;
( \Phi ' + {\Phi \over r}) =   0 ,
$$
$$
f'' +  {4\over r} \; f' - 2 e \; f
\; (2 + e r^{2} K)  \; K -
{\Sigma ' \over  \Sigma} \;  ( f ' + {f \over r}) =   0  ,
$$
$$
K'' + {4 K' \over r} +
 e \; {(f^{2} - \Phi^{2})\;  (1 + e r^{2}K) \over
\Sigma ^{2} } -  e \; K^{2} \; (3 + e r^{2}K) +
$$
$$
+ \;
{\Sigma' \over \Sigma} \;  (K' + {2 K \over r}) = 0  .
\eqno(1.8)
$$

\noindent In (1.8)  we have taken $\lambda = 0$ (the Bogomolny-Prasad-Sommerfield
limit), thus in the following only the situation in  absence  of self-interactions
between components of scalar triplet will be examined.

\subsection*{2. Solutions in flat space}

Now let us turn to    eqs. (1.8) specified for flat Minkowski space model (in Sec. 3
and Sec. 4 we will extend the solving procedure to $H_{3}$ and $S_{3}$ models).
As $\Sigma = 1$ eqs. (1.8) take on the form
$$
\Phi'' +  {4\over r} \; \Phi' - 2 e \;\Phi \;
(2 + e r^{2}K)  \; K  =  0  , \;
f'' +  {4\over r}  \; f' - 2 e \; f
\; (2 + e r^{2} K) \; K =  0  ,
$$
$$
K'' + {4 K' \over r} +
e \;(f^{2} - \Phi^{2})  \; (1 + e r^{2}K)  - e \; K^{2} \; (3 + e r^{2} K)  = 0  .
\eqno(2.1)
$$

\noindent It is known that  the dyon system (2.1) can be solved on the base of
some preliminaries  done for purely monopole system. To the latter there corresponds
the following substitution
$$
\Phi^{a} = x^{a} \; \Phi (r)  ,  \qquad W^{a}_{0} = 0  , \qquad
 W^{a}_{i} = \epsilon _{iab}\;   x^{b} \; K(r)
\eqno(2.2a)
$$

\noindent and respective radial equations are
$$
\Phi'' +  {4\over r} \;  \Phi' - 2 e \; \Phi
(2 + e r^{2}K) \; K  =  0  \;,
$$
$$
K'' + {4 K' \over r} -  e  \; \Phi^{2} \;  (1 + e r^{2}K)
 -  e \; K^{2} \; (3 + e r^{2}K)  = 0  \; .
\eqno(2.2b)
$$

\noindent Turning again to eqs.  (2.1) and setting
 $f = c \; \Phi$ , where $ c $ is a constant, one  comes to
$$
f = c  \; \Phi  , \qquad
{d^{2} \over dr^{2}}\; \Phi  +  {4 \over r}  \; {d \over dr} \Phi -
2e\;  \Phi \; (2 + e r^{2} K)\;  K = 0 \; ,
$$
$$
{1 \over 1 -c^{2}}  \; (
{d^{2} \over dr^{2}} \; K  +  {4 \over r} \; {d \over dr} K )  -
 e\; \Phi^{2}  \;(1 + e r^{2} K)   -
{e K^{2} \over 1 - c^{2} } \; (3 + e r^{2} K ) = 0  .
$$

\noindent   From these,  having introduced a new radial variable
$
r \;\; \rightarrow \;\; (1-c^{2})^{1/4} \; r = \tilde{r}
$
and  a new function  $\tilde{K}$
$$
{K(r) \over \sqrt{1 -c^{2}}}  = \tilde{K}\;  [(1-c^{2})^{1/4}r] \; ,
$$

\noindent one obtains a system of the above type (2.2b).
Therefore the dyon functions have been reduced  to monopole ones:
$$
\Phi =  \tilde{\Phi} \; [(1-c^{2})^{1/4} r]  , \qquad
 f(r) = c  \; \Phi (r)  ,
 $$
 $$
K(r) = \sqrt{1 -c^{2}}\;  \tilde{K} \; [(1-c^{2})^{1/4} r]  \;   .
\eqno(2.3)
$$

\noindent   Taking this in mind further we will examine only the purely
monopole equations. (2.2b).

For further work instead of  $\Phi (r)$  and  $K(r)$ in  (2.2b) it is
convenient to use new functions $f_{1}$ and $f_{2}$:
$$
1  +  e \;r^{2}\; K(r) = r\; f_{1} (r) \; , \qquad
1 +  e \; r^{2}\; \Phi (r) = r \;  f_{2} (r) \; .
\eqno(2.4)
$$

\noindent Correspondingly, eqs.  (2.2b) will assume the form
$$
2  ( f'_{2}  +  f_{1}^{2}  )  +  ( f_{2}''  -
 2 f_{1}^{2} f_{2} ) = 0  ,
 $$
 $$
2 ( f_{1}'  +  f_{1}  f_{2}  )  + r  ( f_{1}''
 -  f_{1}f_{2}^{2}  -  f_{1}^{3}  ) = 0   \;.
\eqno(2.5)
$$

\noindent One can solve these equations by demanding four equalities
$$
f'_{2}  +  f_{1}^{2}   = 0  ,         \qquad
 f_{2}''  -
2 f_{1}^{2} f_{2} = 0  , \;
$$
$$
 f_{1}'  +  f_{1}  f_{2}   = 0 , \qquad
f''_{1} - f_{1} f_{2}^{2} - f_{1}^{3} = 0 .
\eqno(2.6a)
$$

\noindent It is easily verified that second and forth relations are
simple results from first and thirdth ones.
Thus, finally we have only two equations:
$$
f'_{1} = -f_{1} f_{2}  , \qquad  f'_{2} = - f_{1}^{2}  , \qquad
\;\; or  \;\;
$$
$$
f_{2} = - {f_{1}' \over f_{1}}  , \qquad
 ( {f_{1}' \over f_{1}}  )' = f_{1}^{2}  .
\eqno(2.6b)
$$

\noindent The task has reduced to a single differential equation
$$
( {f_{1}' \over f_{1}}  )' = f_{1}^{2} \; .
\eqno(2.6c)
$$

\noindent  Making evident operations one can get
$$
{d \over dr } \; \left [  \; (\ln f_{1})'  \; \right ]^{2} =
{d \over dr}\;  f_{1}^{2} \;  .
$$

\noindent
From this it follows
$$
(\ln f_{1})' = \pm \; \sqrt{f_{1}^{2}  + c_{1}} \; , \qquad
\int {d f_{1} \over f_{1} \; \sqrt{c_{1} + f_{1}^{2}}} =
\pm \;  (r  +  const)  \;  .
$$

\noindent  Depending upon the sign of the constant $c_{1}$
we have three different solutions:
$$
\underline{c_{1} = 0}, \qquad  f_{1} = \pm \; {1 \over r +b }  \; ,
\;  f_{2} = {1 \over r + b} \;  .
\eqno(2.7a)
$$
$$
\underline{c_{1} < 0 }, \qquad    f_{1} = \pm  \; {a \over \sinh (a r + b) } \;   ,
\;
f_{2} = {a  \over \tanh (a r + b)}  .
\eqno(2.7b)
$$
$$
\underline{c_{1} > 0 } , \qquad  f_{1} = \pm  {a \over \sin (a r + b) }  ,
\;
f_{2} = {a  \over \tan (a r + b)} .
\eqno(2.7c)
$$

\noindent Here it should be noted that in going from (2.6b) to (2.6c) we have
missed one simple (but important) solution\footnote{Which is to be
interpreted as abelian Dirac's nonopole being placed into background of
the non-abelian  theory.}
$$
f_{1}(r) = 0    \;\; , \qquad  f_{2} = const \; .
\eqno(2.7d)
$$

\subsection*{3. Some technical details}

In curved space models $H_{3}$ and $S_{3}$, analogously to the flat space $E_{3}$,
there exists possibility to construct dyon functions in terms of purely
monopole's ones  (all details are omitted).  By virtue of this in the
 following
we will examine  only  the  purely monopole ansatz. Instead of
$K(r)$  and  $\Phi (r)$   (See (1.8))  let us  determine
new ones:
$$
1 + e r^{2} K(r)   = A(r)  , \qquad
 e  r^{2} \Phi (r) = B(r)  \;  .
\eqno(3.1)
$$

\noindent
The system (1.8) (setting $f(r) \equiv 0 $) will assume the  form
$$
B'' -  {2 BA^{2} \over r^{2}}  +  {\Sigma' \over \Sigma}
({B \over r } \;- \; B') = 0  , \;
$$
$$
A''  -  { A  B^{2} \over  r^{2} \Sigma^{2} }  +  {A(1 -A^{2}) \over r^{2}}
+ {\Sigma ' \over \Sigma }  A' = 0  .
\eqno(3.2)
$$

\noindent For $A(r)$ and $B(r)$ we will take substitutions
$$
A = c  f_{1}(R)   ,             \qquad
B = a  f_{2}(R)  +  b   .
$$

\noindent
Here  $a,  b,  c, R $ stand for  some yet unknown functions
with respect to  $r$, whereas   $f_{1}$ and  $f_{2}$ are assumed
to obey two relationships (See (2.6b))
$$
{d \over dR } \; f_{1} = - \; f_{1}  \; f_{2}  \; ,
\qquad
{d \over dR } \; f_{2} = - \; f_{1}^{2} \; .
$$

\noindent Therefore they coincide with the above $f_{1}$ and $f_{2}$
(2.7).
Initial equations (1.8) give us the following relationships
for these four functions:
$$
a'' + {\Sigma' \over \Sigma } \; ( { a \over r } - a' )  = 0  \; ,
\qquad
b'' + {\Sigma' \over \Sigma } \; ( { b \over r } - b' )  = 0  \;,
$$
$$
c^{2} = {a^{2} \over  \Sigma^{2} }  , \qquad
(R')^{2} = {a^{2} \over r^{2} \Sigma^{2} }  , \qquad
\;
2 b R'=    -3a' +  2 { \Sigma' \over \Sigma}
 a  +  {a \over r}    \; .
\eqno(3.3)
$$

\noindent
Take notice that the functions $a(r)$ and $b(r)$ satisfy
linear differential
equations. Hence, one can find general expressions for $a(r)$ and $b(r)$,
and further
determine $c(r)$ and $R(r)$, and finally  all these  are to be substituted into
the last equation in (3.3).

\subsection*{4. Monopole in   $H_{3}$-space}

In case of  Lobachevski model,
the equations for  $A(r)$ and $b(r)$ are the same $(a, \; b = g)$:
$$
{d^{2} \over dr^{2}} g  + {r \over 2(1 - r^{2}/4)} {d \over dr}g
- {1 \over 2(1 - r^{2}/4)} g = 0  .
$$

\noindent Two linearly independent solutions are
$   g_{1} = r  , \;  g_{2} = (1 + r^{2} /4)  $.
Therefore
$$
a = a_{1} \;  r  +  a_{2} \; (1+r^{2} /4)  ,
\qquad
b = b_{1} \; r  +  b_{2} \;  (1+r^{2} /4)\;  .
\eqno(4.1)
$$

\noindent Further for $c(r)$  and  $R(r)$ we have
$$
c(r) =   \delta \;  \left   ( \;
 a_{1} \;  { r \over 1 - r^{2}/4} \; +  \; a_{2} \;
{1+r^{2} /4 \over 1 - r^{2}/4} \; \right ) \;   ,
\eqno(4.2a)
$$
$$
R(r) =
\epsilon \; \left ( \;
a_{1} \;  \int { dr \over 1 -r^{2} /4 } \; +  a_{2} \;
\int { 1+r^{2} /4  \over
1 - r^{2} /4 } dr  \right ) =
$$
$$
=  \epsilon\;  (\;   a_{1} \; 2 arc \; tanh {r \over 2} \; + \;a_{2} \;
\ln {r \over 1 - r^{2}/4}
 ) +  const  \; .
\eqno(4.2b)
$$

\noindent  Turning to  the last relation in (3.3) and having for its left side
$$
2ab = 2  \; [ \; a_{2} \; b_{2}  +  (a_{1}\; b_{2} + b_{1}\; a_{2})  r  +
(a_{1}\; b_{1} + {1 \over 2} \; a_{2} \;b_{2}) \; r^{2}  +
$$
$$
+   {1 \over 4} \;
(a_{1} \; b_{2} + a_{2} \; b_{1}) r^{3} +  {1 \over 16} \; a_{2} \; b_{2} r^{4} \; ]
$$

\noindent and for its right side
$$
\epsilon \; r \; \Sigma \;  (- 3a' + 2 {\Sigma \over \Sigma} \;  a + {a \over r}) =
\epsilon \; [\;  a_{2}  -  2 a_{1} \; r  -  {5\over 2} \;  a_{2} \; r^{2}  -
{1 \over 2}\;  a_{1} \; r^{3}  +  {1 \over 16}\; a_{2} \; r^{4}  \; ] \;  ,
$$

\noindent we readily reach at three relations
$$
2 \; a_{2} \; b_{2} = \epsilon \;  a_{2}   , \qquad
  (a_{1} \; b_{2} + a_{2}\; b_{1}) = - \epsilon  \; a_{1}  ,
  $$
  $$
  ( a_{1} \; b_{1} + {1\over 2} \; b_{2} \; a_{1}) = - \epsilon  \; {5\over 4} \;  a_{2} \;  .
\eqno(4.3)
$$

Supposing $a_{2} \neq 0$, from first relation it follows
$b_{2} = \epsilon /2$, and two remaining ones take on the form
$$
a_{2} \; b_{1} = - {3 \over 2}\; \epsilon  a_{1}  , \qquad
a_{1} \; b_{1}  = -  {3\over 2}\; \epsilon a_{2}\; .
$$

\noindent  Further getting
$$
{a_{2} \over a_{1}} = +  {a_{1} \over a_{2}}\;\;  \Longrightarrow \;\;
({a_{2} \over a_{1}} )^{2} = + 1 \;  ,
$$

\noindent and therefore
$$
{a_{2} \over a_{1}}  =   \pm \;   1  , \qquad
b_{1} = \mp \;  {3 \over 2}  \; \epsilon \;  .
$$

\noindent So  at $a_{2} \neq 0$ we arrive at  the solution
$$
{ \bf I. } \qquad
a = a_{1} \;  [ \;  \;  \pm \;   (1 + {r^{2} \over 4 }) \; ] \; ,
\qquad
b = {\epsilon \over 2}   \; [\;   \mp\;  3 r  +
(1 + {r^{2} \over 4 })\; ] \;   ,
$$
$$
c(r) =   \delta \;    a_{1} \;   { r\;  \pm \;  (1 + r^{2}/4) \over  1 - r^{2}/4} ) \; ,
$$
$$
R(r) =
\epsilon \;  a_{1} \;
\left (\;  2 \; arc \; tanh {r \over 2} \;\pm \;
a_{2} \; \ln {r \over 1 - r^{2}/4}  \;\right  )\;  + \;  const \;  .
\eqno(4.4)
$$

There are other possibilities  too.
Thus,  having supposed  $a_{2} = 0$, from (4.3) it follows
$$
a_{1}  \; b_{2} = - \epsilon \; a_{1}  , \qquad
a_{1} \;  b_{1} = 0 \; .
$$

\noindent From this it follows:
$ if \; a_{2} =0  ,  \;\; a_{1} \neq 0 \;\; $  then $
b_{1} = 0  ,  \;\;  b_{2} = - \epsilon   $.
Therefore
$$
{\bf II \; } : \qquad
 a(r) = a_{1}\;  r\;  ,
\qquad
b(r) = - \epsilon \;  (1 + {r^{2} \over 4})\; ,
$$
$$
c(r) = \delta  \; {a_{1}  r \over 1 - r^{2}/4} , \qquad
R(r) = \epsilon \;  a_{1} \;  (\; 2 \; arc \; tanh {r \over 2} \;) \;  .
\eqno(4.5)
$$

\noindent
And third solution there exists at $a_{2}, a_{1} =0 $:
$$
{\bf III.} \qquad
a(r) = 0 ,
\qquad
b(r) = b_{1}  \; r  +  b_{2} \; (1 + {r^{2} \over 4 })\; ,
$$
$$
c(r) = 0  , \qquad  R(r) = const  .
\eqno(4.6)
$$

\subsection*{5. Monopole in $S_{3}$-space. }

Analysis for $S_{3}$-model can  be accomplished   in analogous way. Omitting
all details we write down final results:
$$
{\bf I.} \qquad
a = a_{1}  \; [\;  r \;  \pm i  \;  (1 - r^{2} /4) \; ] \; ,
\;
$$
$$
b = {\epsilon \over 2} \;  [\;  \pm 3 i \; r +   (1 - r^{2} /4) \; ]\;   ,
\qquad
c = \delta\;  { a(r) \over 1 + r^{2}/4}\;  ,
$$
$$
R = \epsilon  \; a_{1}  \;\left ( \; 2 \; arc \; tan {r \over 2}\;  \pm \;
 i \; \ln {r/2 \over 1 + r^{2}/4 } \; \right  )  +  const  \;   ;
\eqno(5.1)
$$
$$
{\bf II.} \qquad
a = a_{1}\;  r \; ,
\;
b(r) = - \epsilon  \; (1 - r^{2} /4 ) \; ,
$$
$$
c = \delta \;  {a_{1} r \over 1 + r^{2}/4}  \; ,
\;
R = \epsilon 2 \; arc \; tan  {r \over 2} + const \; ;
\eqno(5.2)
$$
$$
{\bf III. } \qquad
a = 0 \;   ,
\;
b =  b_{1} \;  r  +  b_{2} \; (1 - r^{2} /4 )\;  ,
$$
$$
\;
c(r) = 0  \; ,\;  R(r) = const \; .
\eqno(5.3)
$$

\subsection*{6. Discussion}

The solution of the type I in $S_{3}$-model is complex and it can be withdrawn
from physical consideration  because of the initial requirement
of reality of all  fields.  The I-type solution in $H_{3}$ space,
though being real,
should be  regarded as  non-physical because  it is a direct counterpart of
complex that in  $S_{3}$ space. The solutions of the  type III in both
$H_{3}$ and $S_{3}$  models can be evidently qualified as degenerated and
of small physical  interest.
Thus, only the type II solutions in both spaces are to be of physical meaning ant they
must be analyzed further.

And a final remark. Among all solutions in three spaces models $E_{3}, H_{3},S_{3}$
there exist just three  ones
which reasonably and in a one-to-one correspondence can be associated
with respective geometries. The situation can be characterized by the schema
$$
\left. \begin{array}{lccc}
  &  E_{3}  &  H_{3}  &  S_{3}  \\
(ar+b)  &  *  &  -   &  -  \\
\sinh (ar+b)  & - &  *  & -  \\
\sin (ar+b)  &  -  & -  &  *
\end{array}  \right.
$$

It should be noted that the known non-singular BPS-solution in the flat
Minkowski space  can be understood as a result of somewhat artificial
combining the Minkowski space background with a possibility naturally
linked up with the Lobachewski geometry
$$
\left. \begin{array}{lccc}
  &  E_{3}  &  H_{3}  &  S_{3}  \\
(ar+b)  &  -  &  -   &  -  \\
\sinh (ar+b)  & * &  -  & -  \\
\sin (ar+b)  &  -  & -  &  -
\end{array}  \right.
$$

The standpoint is brought forth
that  of primary interest should be regarded only three specifically  distinctive
solutions --- one for every curved space background. In the framework of those
arguments  the generally accepted status of the known monopole BPS-solution
should be critically reconsidered and even might be given away.

\begin{center}
References
\end{center}

1.
G. 't Hooft
{\em Monopoles in  unified gauge theories.}
Nucl. Phys. B. 1974. Vol. 79, No 2. P. 276-284.

\noindent
2.
B. Julia, A.  Zee.
{\em Poles with both magnetic and electric charges in non-Abelian gauge theory.}
Phys. Rev. D.  1975.  Vol. 11, No 8. P. 2227-2232.

\noindent
3.
M.K. Prasad, C.M.  Sommerfield.
{\em Exact classical solution of the 't Hooft monopole
and Julia-Zee dyon.}
Phys. Rev. Lett. 1975.  Vol. 35, No 12. P. 760-762.

\noindent
4.
F.A. Bais,  R.J. Russel.
{\em Magnetic-monopole solution of non-Abelian gauge theory in
curved space-time.}
 Phys. Rev. D. 1975. Vol. 11, No. 10.  P. 2692-2695.

\noindent
5.
E.B Bogomolny.
{\em Stability of classical solutions} (in Rissian).
Yad. Physika. 1976.  Vol. 24, No 4. P. 861-870.

\noindent
6.
P. Van Nieuwenhuizen,  D. Wilkinson, M. Perry.
{\em Regular solution  of 't Hooft's magnetic monopole in  curved space.}
Phys. Rev. D.  1976.  Vol. 13, No 4.  P. 778-784.

\noindent
7.
D. Wilkinson, F.A. Bais.
 Phys. Rev. D.  1979. Vol. 19. P. 2410-2415.

\noindent
8.
M. Komata, M. Kasuya.
{\em An exact Schwinger dyon solution in the Kerr-Newman space-time.}
Nuovo Cim. A.  1981. Vol. 66, No 1.  P. 67-70.

\noindent
9.
M. Kasuya.
{\em An exact rotating Julia-Zee dyon solution with the Kerr-Newman metric.}
Phys. Lett. B. 1981. Vol. 103, No 4-5.  P. 353-354.

\noindent
10.
M. Komata. {\em Comments on an exact rotating Julia-Zee dyon solution
with the Kerr-Newman metric.}
Phys. Lett. B.  1981.  Vol. 107,  No.  1-2. P. 44-46.

\noindent
11.
V.K. Schigolev. {\em Magnetic monopole in general relativity.}(in Russian).
Izvest. Wuzow. Physika. 1982, No  11. P. 89.

\noindent
12.
V.K. Schigolev.
{\em Magnetic monopole solutions of the non-abelian theory
in the expending universe} (in Rissian).
 Izvest. Wuzow. Physika.  1984. Vol. 27. No 8.  P. 67-71.

\noindent
13.
V.N. Melnikov, V.K. Shigolev.
{\em Exact S0(3) magnetic monopole solutions in the expanding universe.}
Lett. Nuovo Cim.  1984.  Vol. 39, No 5.  P. 364-368.

\noindent
14.
J. Villarroel. {\em  Yang-Mills solutions in $S^{3}\times S^{1}$.}
J. Math. Phys.  1987. Vol. 28, No 11. P. 2610-2613.

\noindent
15.
B.S. Pajput, G. Rashmi
{|em Unification of fields associated with dyons. Gravito-dyons in non-Abelian
gauge theory.}  Indian J. Pure and Appl. Phys.
1989.  Vol. 27. No 1. P. 1-9.

\noindent
16.
A.A. Ershov, D.V. Gal'tsov.
{\em Non-existence of regular monopoles and dyons in the SU(2)
Einstein-Yang-Mills theory.}
Phys. Lett. A.  1990.  Vol. 150.  No  3-4. P. 159-162.

\noindent
17.
Y. Yang.  {\em On the BPS limit in the  classical SU(2) gauge theory.}
J. Phys. A.  1990.  Vol. 23. No 9.  L. 403-407.

\end{document}